# Uncovering dynamical equations of stochastic decision models using data-driven SINDy algorithm

Brendan Lenfesty[1], Saugat Bhattacharyya[1], KongFatt Wong-Lin[1,*]

[1]Intelligent Systems Research Centre, School of Computing, Engineering and Intelligent Systems, Ulster University, Magee campus, Derry~Londonderry, Northern Ireland, UK

[*]Corresponding author: KongFatt Wong-Lin (k.wong-lin@ulster.ac.uk)

# Abstract

Decision formation in perceptual decision-making involves sensory evidence accumulation instantiated by the temporal integration of an internal decision variable towards some decision criterion or threshold, as described by sequential sampling theoretical models. The decision variable can be represented in the form of experimentally observable neural activities. Hence, elucidating the appropriate theoretical model becomes crucial to understanding the mechanisms underlying perceptual decision formation. Existing computational methods are limited to either fitting of choice behavioural data or linear model estimation from neural activity data. In this work, we made use of sparse identification of nonlinear dynamics (SINDy), a data-driven approach, to elucidate the deterministic linear and nonlinear components of often-used stochastic decision models within reaction time task paradigms. Based on the simulated decision variable activities of the models, SINDy, enhanced with approaches using multiple trials, could readily estimate the dynamical equations, choice accuracy and decision time of the models across a range of signal-to-noise ratio values. In particular, SINDy performed the best using the more memory intensive multi-trial approach while trial-averaging of parameters performed more moderately. The single-trial approach, although expectedly not performing as well, may be useful for real-time modelling. Taken together, our work offers alternative approaches for SINDy to uncover the dynamics in perceptual decision-making, and more generally, for first-passage time problems.

## Introduction

Decision-making is the keystone to unlocking higher cognitive functions (Dreher & Tremblay, 2016). More specifically, perceptual decision-making involves transforming sensory information into motor actions, in which the latter overtly provides choice information such as choice accuracy and reaction time (Luce, 1986; Roitman & Shadlen, 2002). Its neural correlates have also gradually been revealed (Gold & Shadlen, 2007; Dreher & Tremblay, 2016; Hanks & Summerfield, 2017; Najafi & Churchland, 2018; O'Connell et al., 2018; O'Connell & Kelly, 2021).

Previous studies have indicated that perceptual decision-making process can be described by the sequential sampling modelling framework in which noisy sensory evidence is accumulated over time until reaching some internal decision threshold upon which a choice will be committed (Gold & Shadlen, 2007; O'Connell et al., 2018). Popular cognitive computational models of perceptual decision-making include the drift-diffusion model (DDM) and its variants (e.g. Ratcliff, 1978; Ditterich, 2006; Ratcliff et al., 2016; Asadpour et al., 2024). Connectionist or neural network representation of the sequential sampling models include the leaky competing accumulator (LCA) model (Usher & McClelland, 2001) which was shown to approximate the classic DDM process (LCA-DDM) under finely tuned conditions (Bogacz et al., 2006). In reaction-time tasks, the decision formation process can be mathematically described as a first-passage time problem (Gillespie, 1992; Bogacz et al., 2006; Kampen, 2007; Shinn et al., 2020). Although simple to implement, to understand and mathematically tractable, these models' variables and parameters may not directly relate to neurophysiology, and may account for limited neural data (e.g. Ditterich, 2006; Gold & Shadlen, 2007; Kelly & O'Connell, 2013).

More biologically based models of perceptual decision-making have been developed to bridge from neural to behavioural data while providing additional model constraints (Wang, 2002; Wong & Wang, 2006; Roxin & Ledberg, 2008) O'Connell et al., 2018). Models informed by neural activities can better delineate underlying decision processes and reduce model mimicry based only on choice behaviour (Ratcliff and Smith, 2004; O'Connell et al., 2018; Bose et al., 2020). These models are often accompanied by intrinsic nonlinear behaviours which may allow better replication of neural activity profile while accounting for choice behaviour (e.g. Niyogi & Wong-Lin, 2013). Perhaps the simplest nonlinear dynamical model for two-choice tasks can be mathematically described by a normal form with pitchfork bifurcation, a nonlinear bistable (NLB) model (Zhou et al., 2009), which resembles biologically grounded mean-field models (Wong & Wang, 2006; Roxin & Ledberg, 2008).

Several model fitting algorithms had been used to identify decision model parameters to account for behavioural data. For decision network modelling, model parameters were either fitted to neural (Friston et al., 2003) or behavioural data (e.g. Bogacz & Cohen, 2004; Pinto et al., 2019; Shinn et al., 2020), or did not employ automated learning algorithms (e.g. Wong & Wang, 2006; Niyogi & Wong-Lin, 2013). More complex and high-dimensional recurrent neural network models have also been trained on choice behavioural data before comparing their neural activities to those of recorded neurons (e.g. Mante et al., 2013). However, there are limited studies that link these high-dimensional models to lower-dimensional models (e.g. Roach et al., 2023). Importantly, there is limited data-driven modelling work that automatically and directly uncovers low-dimensional nonlinear dynamical equations underlying decision formation from internally noisy decision variable data, which in turn predicts choice behaviour.

In this work, we address the above by adopting a data-driven approach for elucidating the governing dynamical equations of stochastic decision models. Some existing data-driven methods are sensitive to noise, while others are computationally expensive (Raissi et al., 2019), or limited to specific fields (Schmid, 2010). Other methods are "black boxes" that do not provide explicit equations or explanations (Lipton, 2018). Importantly, none have been tested on stochastic decision dynamical models in reaction-time tasks, which can be mathematically described as first-passage time problems.

Here, we will use the sparse identification of nonlinear dynamics (SINDy) (Brunton et al., 2016; de Silva et al., 2020) to identify parsimonious linear or nonlinear equations of low-dimensional single-trial decision dynamics, given existing evidence suggesting decision neural dynamics are embedded in low-dimensional space (Ganguli et al., 2008; Shenoy & Kao, 2021; Steinemann et al., 2023). The SINDy algorithm will be evaluated on simulated noisy data generated from often-used decision models, namely, DDM, linear LCA, LCA-DDM and NLB. Decision variable activity, choice behaviour and parameters of the original models at single trials and across trials will be compared with that estimated by SINDy. The focus will be on identifying the deterministic components of these models. In addition to the standard single-trial SINDy approach, we also make use of approaches using multiple trials to enhance SINDy's performance. Our general findings show that SINDy can readily elucidate the dynamical equations and account for decision models' variable profile and choice behaviour within reaction-time tasks.

## Methods

### 2.1 Model description

Four sequential sampling two-choice decision-making models were simulated, and their decision variable activities and choice behaviours were compared with the respective estimated models elucidated by SINDy. Specifically, the standard DDM (Ratcliff, 1978), the linear version of LCA model (Usher & McClelland, 2001), the LCA model which approximates the DDM process (LCA-DDM) (Usher & McClelland, 2001; Bogacz et al., 2006), and the NLB model (Zhou et al., 2009) were used.

The two-choice standard DDM can be described by a 1-dimensional stochastic differential equation (Ratcliff et al., 2016):

$$dX = A\,dt + \sqrt{dt}\,\sigma\,\eta \tag{1}$$

where $X$ denotes some internal decision variable, $A$ is the drift rate determined by the input signal or stimulus, $\eta$ is a random variable that follows a Gaussian distribution with mean of zero and standard deviation of one, $\sigma$ is the noise size, and $t$ is time with time step $dt$. Values of $A$ were varied between $0$ and $0.04$ (to mimic varying choice task difficulty) while $\sigma$ was fixed at $0.11$.

During decision formation, $X$ started with an initial value of $0$ and was integrated over time via Equation (1) such that it reaches either a prescribed upper or lower decision threshold, indicating one of the two choices being made. The upper threshold for a correct choice (for positive drift rates) was set at 1, while the lower threshold for an error choice was $-1$. Once a threshold has been reached, the integration process is ceased (i.e. an absorbing threshold), and this time from stimulus onset is defined as the decision time.

The linear LCA model for 2-choice task can be described by two coupled stochastic differential equations (Usher & McClelland, 2001):

$$dy_1 = (-ky_1 - by_2 + S_1)dt + \sqrt{dt}\,\sigma\,\eta_1 \tag{2}$$

$$dy_2 = (-ky_2 - by_1 + S_2)dt + \sqrt{dt}\,\sigma\,\eta_2 \tag{3}$$

where the $y_i$'s are the decision variables for choice $i$, $k$ and $b$ are the decay rate and inhibitory coupling constants while $S_i$'s are the stimulus inputs. $\sigma$ is the noise size and $\eta_i$ is a random variable following a Gaussian distribution with mean zero and standard deviation of one. Unlike the DDM, there is only one decision for each $y_i$, and when one of the two $y_i$'s reaches its threshold first, the choice $i$ is made and the process is ceased.

With appropriately configured parameter values, the linear LCA can lead to runaway ramping (acceleration) over time for one of the decision variables due to the existence of a metastable saddle steady state (Usher & McClelland, 2001); or it may converge towards some stable steady state (a fixed-point attractor) (Usher & McClelland, 2001). We shall focus on the metastable version of the linear LCA as the stable steady state version will be investigated using the NLB model (see below).

Importantly, the DDM process can be approximated from the linear LCA model when the decay rate term $(-ky_i)$ and mutual inhibition $(-by_j)$ are equal and high, i.e. finely tuned parameters (Bogacz et al., 2006). For the linear LCA model the mutual inhibitory factor $b$ and decay rate $k$ were set at 4 and 3, respectively, while $S_1$ was varied between 1.85 and 3.05 and $S_2$ fixed at 1.85. For the LCA-DDM model, $b$ and $k$ were both 10, while $S_1$ was varied between 3 and 3.04 and $S_2$ fixed at 3. Both models had noise $\sigma$ at 0.11.

The NLB model is described by the following stochastic differential equation (Zhou et al., 2009):

$$\tau_X \, dX = [\varepsilon X + X^3 - X^5 + b_0]dt + \sqrt{dt} \, \sigma \eta \qquad (4)$$

where $X$ is the decision variable, $b_0$ and $\varepsilon$ represent the biased (signal) and non-biased stimulus inputs, respectively, while $\sigma$ is the noise size and $\eta$ is a random variable following a Gaussian distribution with mean zero and standard deviation of one. $\tau_X$ is some characteristic time constant of the system. Two decision thresholds were placed equidistant from the starting point. As in the DDM, when either decision threshold was reached, the temporal integration of the decision variable was ceased, and a choice was made. $b_0$ varied between 0 and 0.004, and $\varepsilon$ had a constant value of 0.05. Noise size $\sigma$ of 0.01 was used.

## 2.2 Model simulations

A wide range of values of the signal-to-noise ratios (SNR) for each model were used. The appropriate number of trials for each model is based off the results of confidence intervals of the predicted choice behaviour meeting the 90% confidence interval criterion. Each set of model parameters are generated over 10000 trials. To compare across models, we normalised the decision times with min-max normalisation for both the decision and derived SINDy models.

To numerically integrate the stochastic differential equations in all the models, we employed the Euler-Maruyama method (Higham, 2001). For DDM, a time step of 0.1 a.u. was used, while it was 0.01 a.u. for LCA, LCA-DDM and NLB models. Smaller time steps did not affect the results.

## 2.3 Sparse identification of nonlinear dynamics

The SINDy method makes use of compressed sensing for signal detection to uncover the mathematical equations governing dynamical systems by estimating their model parameters based on the observed system's trajectory in state space (Brunton et al., 2016; de Silva et al., 2020). Specifically, suppose we have a general dynamical system that can be described by $\dot{X} = \frac{dX}{dt} = f(X(t))$ where $X(t) \in \mathbb{R}^n$ is some $n$-dimensional dynamical state vector, $f$ a nonlinear function defining the equations of motion, and the state variables $X$ and $\dot{X}$ in matrix form. We can then define a library $\Theta(X)$ of nonlinear candidate

functions of the $X$ columns which can include constants, polynomials, and other mathematical functions:

$$\boldsymbol{\Theta}(X) = \begin{bmatrix} | & | & | & & | & | & \\ 1 & X & X^2 & ... & \sin(X) & \cos(X) & ... \\ | & | & | & & | & | & \end{bmatrix} \quad (5)$$

and a vector of parameters or coefficients $\boldsymbol{\Xi} = [\xi_1\ \xi_2\ ...\ \xi_n]$ that find the active terms in $\boldsymbol{f}(X(t))$.

With this setup, SINDy can be implemented in three general steps. Firstly, numerical differentiation of the (decision) dynamical variable(s) $X$ is performed to obtain $\dot{X}$. Secondly, feature library is used to find a sparse linear combination that can recreate the dynamics of the system for the appropriate candidate function $\boldsymbol{\Theta}$ to estimate $\dot{X} \approx \boldsymbol{\Theta}(X)\,\boldsymbol{\Xi}$. Thirdly, a sparse regression algorithm is used to determine the active coefficients of $\boldsymbol{\Xi}$, assuming that $\boldsymbol{f}(X(t))$ admits a sparse representation in $\boldsymbol{\Theta}(X)$. Thus, a parsimonious model of the system can be obtained via least squares regression with sparsity-promoting $L_1$ regularisation with parameter $\lambda$:

$$\xi_k = \arg\min_{\xi_k'} \left\| \dot{X}_k - \boldsymbol{\Theta}(X)\xi_k' \right\|_2 + \lambda \left\| \xi_k' \right\|_1. \quad (6)$$

Elucidating the sparse $\xi_k$ will enable replicating the original dynamics, the $k$ index denotes the active coefficients for the sparse vector making up $\boldsymbol{\Xi}$. Thus, given some state trajectory, one can consider SINDy to optimise the minimal combinations of "basis" functions that can replicate the observable trajectory.

Here, we used the Python version of SINDy (PySINDy) (de Silva et al., 2020). We only used polynomials as $f$ in all the considered decision models can be described by a combination of polynomials. We used the Savitsky-Golay filter (Press & Teukolsky, 1990) in PySINDy to smooth out noise. As decision models entail more noise than in previous SINDy studies, we also performed approaches that made use of multiple trials. Specifically, to handle noise effects, we used the multi-trial approach available in SINDy (Brunton et al., 2016), which estimates a model based on trajectories from multiple trials. We also employed a less computationally memory intensive trial-averaging method of model parameters deduced from single trials. It should be noted that SINDy only elucidates the deterministic part of a dynamical system.

For each SNR and model, the SINDy-recreated dynamics will be compared to that of the original model in terms of: (i) single-trial decision variable dynamics and predicted choice accuracy and mean decision time; and (ii) across-trial decision variable dynamics, choice accuracy and mean decision time. We used the same noise generalisation, that is the same random seed, for both the original model and estimated model for both single-trial, across-trial and multi-trial comparisons. This allowed us to clamp the noise factor to understand the effects purely due to the deterministic terms in the models. These steps are summarised in Figure 1.

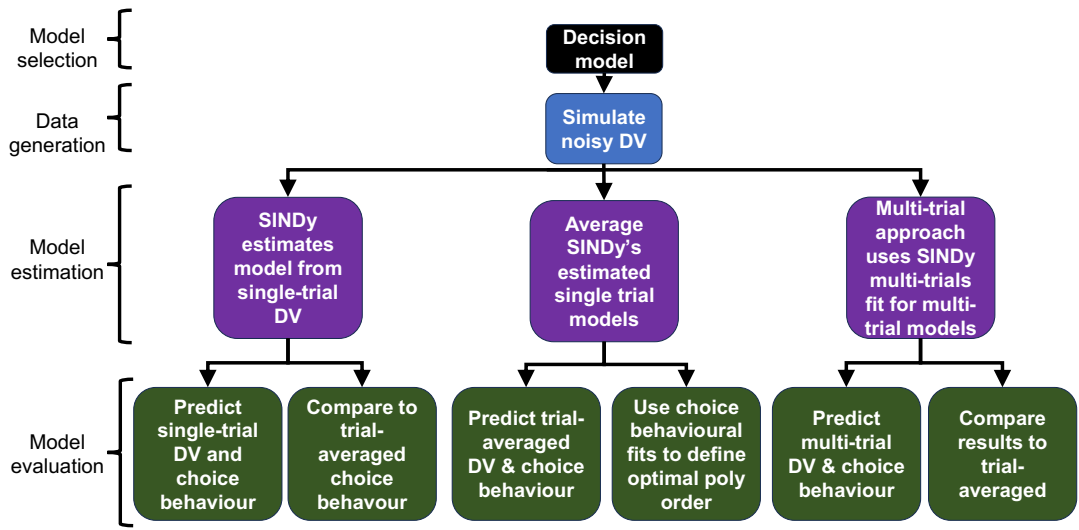

**Figure 1:** Schematic of workflow for SINDy processing and comparison with original models. DV: decision variable.

When using SINDy, it is important to set the polynomial order, as this dictates the complexity of the estimated model. To identify the best polynomial order, we ran simulations with various polynomial orders on the trial-averaged parameters and multi-trial approaches to determine the optimal order that best fits the choice behaviour. For multi-trial DDM, LCA-DDM and LCA SINDy parameter estimation was performed using all 10000 time courses for all 41 SNRs, however for NLB, due to the memory intensive operations resulting from the timestep and SINDy fitting, we could only utilise 4000 of the time courses for parameter estimation.

It is worth noting that fitting each model by trial-averaging the time course of truncated decision variable will lead to an artefact that may seemingly look like stable steady state (fixed point) lying beneath decision thresholds even if the considered models are perfect neural integrators such as DDM or have actual stable steady state beyond the decision thresholds (as in NLB). This approach leads to confusion and poor performance by SINDy. To evade this, one may consider applying SINDy only on earlier time epoch of the trial-averaged decision variable; however, this would again lead to low SINDy performance with the less data available for SINDy. Hence, we did not pursue along these directions.

## 2.4 Statistical analysis

To assess the choice behaviour of simulated decision models versus SINDy's derived decision models, we employed the Kolmogorov-Smirnov (KS) test (Smirnov, 1948). We utilised Cliff's Delta (Cliff, 1993) to quantify the effect size of the differences between choice behaviour distributions.

For both single-trial condition and conditions over multiple trials, we employed a non-parametric bootstrap resampling technique approach (Davison & Hinkley, 1997) to calculate confidence intervals, using a predefined number of bootstrap samples (1000) and aiming for a confidence level of $90\%$. We derived the lower and upper bounds of the confidence intervals based on the $10$th and $90$th percentiles of the bootstrap mean distributions (Efron & Tibshirani, 1994).

To compare the averaged decision variable activities between the original models and the corresponding SINDy-derived models, we split the SNR into 4 different values. First, we identified the high SNR based on two conditions – when the choice accuracy for both the original and SINDy-derived models attained $\geq 95\%$ and their differences are $\leq 1\%$. Then, we equally divide between this SNR value and zero SNR into three equal intervals, leading to categorisation of low and medium SNR values.

To quantify the differences between the model parameters obtained by SINDy and those of the corresponding original models, we compute room mean squared error (RMSE) of the parameters for each condition per trial and then aggregate the values (Supplementary Figure S3).

**2.5 Software and hardware**

The simulations and analyses were run using Jupyter Notebook running on Python 3. We used a Windows machine with four memory cores, Intel i7-7300HQ, and 32GB RAM and the Northern Ireland High Performance Computing (NI-HPC) facility (www.ni-hpc.ac.uk). Source codes will be openly made available upon publication.

# Results

We first investigate how SINDy can estimate single-trial decision variable dynamics simulated from the considered stochastic decision models in reaction-time task. We will then find out how well the single-trial estimated dynamics can predict the aggregated choices for specific SNR. This is followed by SINDy approaches using multiple trials.

## 3.1 Mixed results of choice prediction from SINDY's single-trial parameter estimation

Figures 2A-D (left, blue) show sample simulated decision-variable dynamics with specific SNR for the different decision models. Plotted alongside were the corresponding SINDy's re-created dynamics (Figure 2A-D, left, orange) based on estimation of model parameters of the same trial and using the same random seed as the original model (Figure 1, top). The optimal polynomial order for each trial-averaged or multi-trial model was selected based on the best fit to choice behaviour (choice accuracy and decision time) (Figure S1 and S2). As LCA-DDM was originally a model of first order polynomial, we first consider this order before investigating the optimal zero order.

The presented samples here had the re-created dynamics reaching the same decision threshold as that of the original models, i.e. correctly predicting trials. There were also trials where they reached different decision threshold, i.e.

incorrectly predicted trials (not shown). SINDy's re-created dynamics readily replicated the original time course well, especially at the beginning, before diverging and hence reaching the decision threshold at a different time point (i.e. different decision/first-passage time) than that of the original model, despite having the same seeded noise. This suggested that SINDy might estimate the model parameters slightly differently from that of the original models. This was indeed the case for DDM in Figure 2A (see Supplementary Note 2 for details). For the LCA-DDM, assuming the underlying model order to be of polynomial one, we found that the SINDy-derived model was not as asymmetrical as the original LCA-DDM (Supplementary Note 2). SINDy's derived model of the linear LCA provided a closer approximation and hence better prediction of the dynamics than the LCA-DDM derived model (Supplementary Note 2). Finally, the NLB replicated the activity well despite not capturing the original model parameters as closely as for the LCA (Supplementary Note 2).

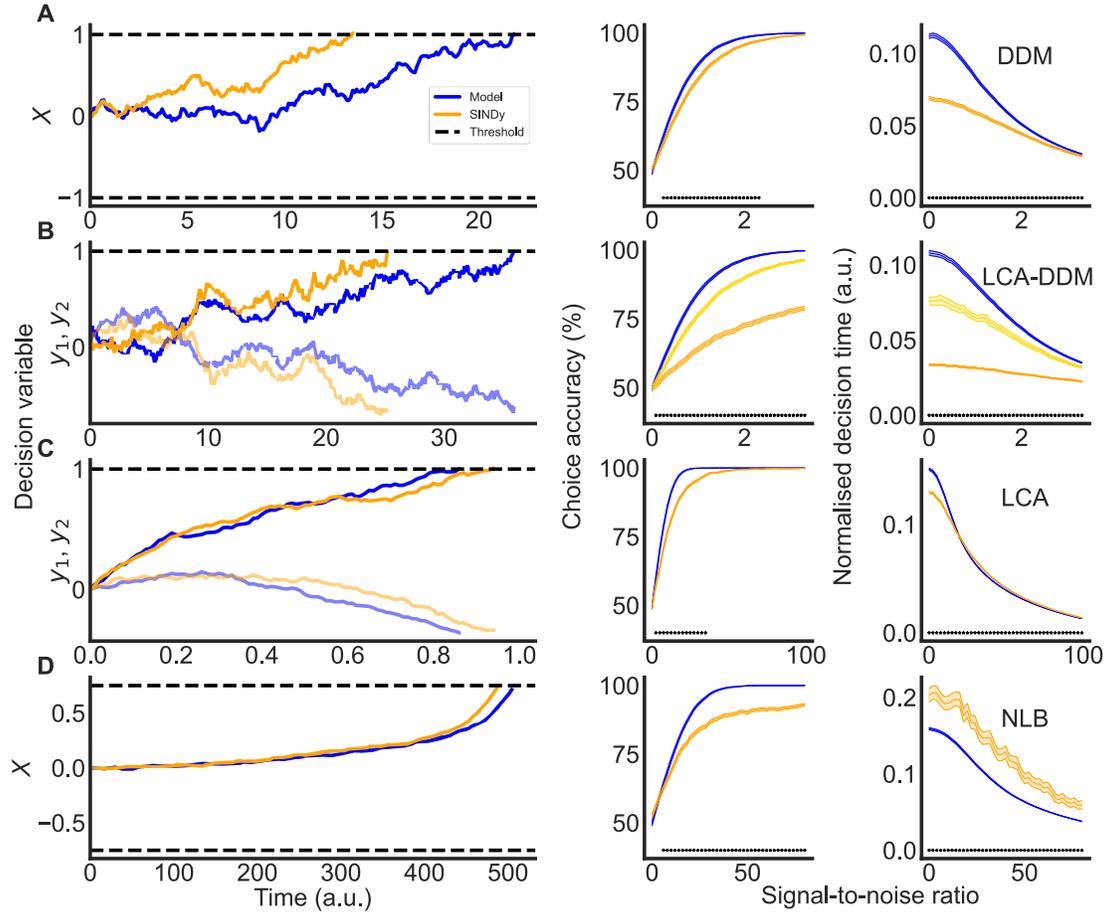

**Figure 2:** Single-trial parameter estimation by SINDy produced mixed results. (A-D) Models from top to bottom: DDM, LCA-DDM, LCA and NLB. Same within-trial random seed used for original and estimated data. Reaction time task (i.e. first-passage time process). Blue (orange): original (estimated) model's data. Left: SINDy-generated decision variable dynamics compared to that of original models. Time from stimulus onset. Dashed: decision threshold. (B-C, left) Faded: Activity of losing competing units. Middle: Choice accuracy. Right: Normalised decision time. Middle, right: Colour label as in left column; (B,

middle, left): gold for the optimal zero order SINDy-derived model. Confidence interval (90%) shown. Asterisks at bottom: statistical significance ($p < 0.05$) between original and derived models.

We next aggregated the trials for specific SNR. From this, we compared the choice accuracy and decision time between the original models and the corresponding models by SINDy. We found that as SNR increased from zero value, the choice accuracy for DDM, LCA-DDM and NLB deviated from that of SINDy's models (Figure 2B, middle, compare blue with orange). Higher SNR generally led to faster decisions, and hence, less generated data revealed to SINDy. This results in higher difficulty in elucidating the model parameters well. However, there are some nuances. First, for DDM, sufficiently high SNR led to better SINDy prediction (Figure 2A, middle, right). Second, SINDy predicted LCA better with higher SNR, and attained the best prediction amongst the models (Figure 2C, middle, right). This was due to the more unstable (runaway) dynamics of LCA, such that noise became relatively less important towards contributing to the overall dynamics. Third, SINDy did not predict LCA-DDM choice behaviour generally well when first-order polynomial was used (Figure 2B orange, middle, right). However, when zero-order polynomial was used, the prediction improved substantially (Figure 2B, gold). This is not surprising, given that the LCA-DDM behaves very similar to DDM through model parameter fine tuning (Bogacz et al., 2006). Interestingly, SINDy predicted readily well for NLB despite its multi-stable states and the presence of noise.

Overall, SINDy predicted noisy single-trial decision variable dynamics and choice behaviour with mixed results, performing well only for certain models and SNR ranges. We next investigate whether trial-averaged estimated model parameters and multi-trial SINDy can improve the prediction of the averaged decision variable dynamics and choice behaviour.

## 3.2 Trial-averaged and multi-trial parameter estimation enhanced choice prediction for SINDy

For the trial averaging of parameters approach, we took SINDy's estimated parameters across trials and then used these averaged parameters, but with the multi-trial approach we fit all trials for each signal to noise ratio and used the generated models' parameters. Each was used to re-create the decision variable dynamics choice behaviour (Figure 1). Figure 3 (compared to Figure 2) shows smaller confidence intervals for the choice behaviour as compared to that for single-trial choice behaviour. Note that for the LCA-DDM (Figure 3B), only the zero-order polynomial was shown as it provided the best fit to choice behaviour for the trial-averaged and multi-trial approaches, as compared to the original order of polynomial one (Figure S1). Importantly, the SINDy predictions of choice behaviour were improved, further supporting trial-averaged and multi-trial approaches.

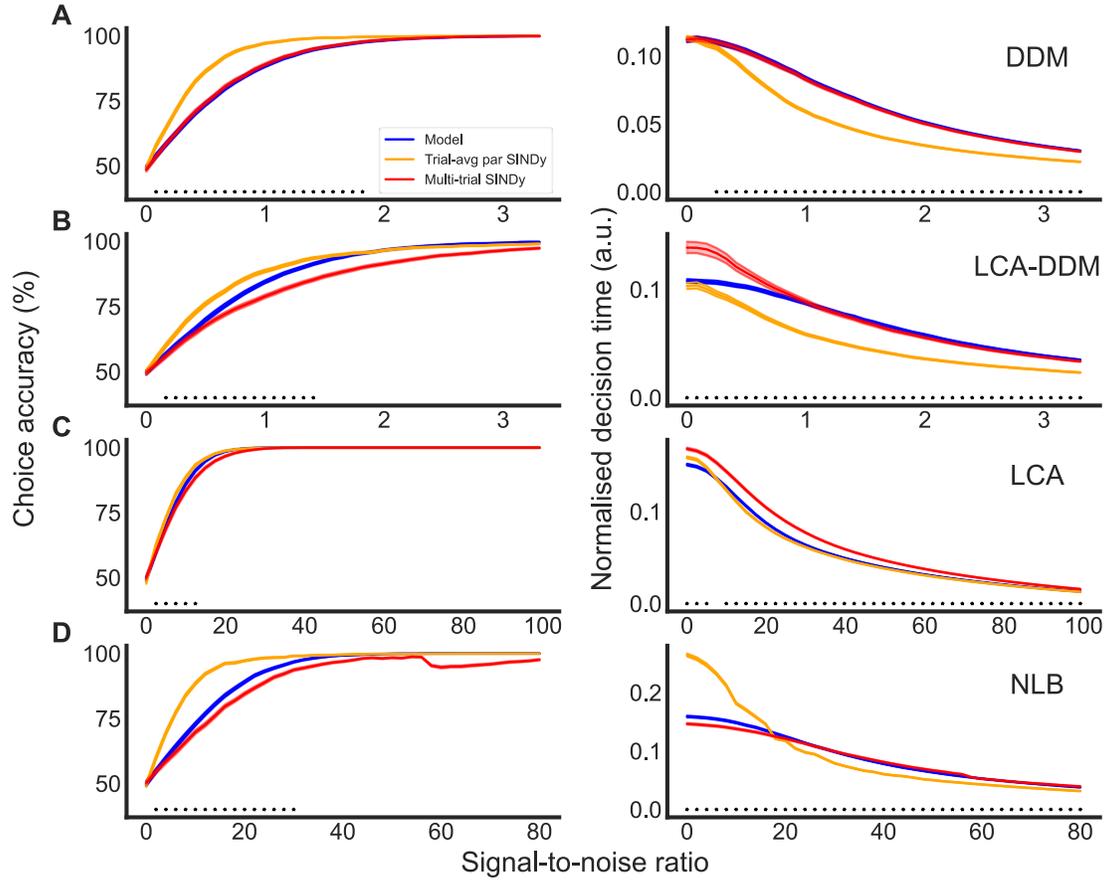

**Figure 3:** Improved SINDy-derived model predictions of choice behaviour with trial-averaging and multi-trial approaches. (B) Only with model with zero order polynomial. Labels as in Figure 2. Choice behaviour shown only for the optimal polynomial order.

Next for the trial-averaged of parameters approach, we sorted the SNRs into zero, low, medium and high values (see Methods) and averaged the decision variable activities regardless of choice. For the DDM, higher SNRs could replicate the decision dynamics well, but not as much for low and zero SNRs (Figure 4A), consistent with SINDy's predicted choice (Figure 3A). The finely tuned LCA-DDM parameters also struggled to be replicated by SINDy even with the trial averaging approach; the averaged LCA-DDM activity for low and zero SNR (Figure 4B, two left columns) struggled to predict the decision time as decisions occurred much earlier than the LCA-DDM (Figure 3B, right column). Despite this, there was a slight improvement for decision time than that using single trials, but much more so for choice accuracy. The fine-tuned parameters meant that any slight difference in predicted values between its inhibitory term and decay rate by SINDy might lead to very different dynamics, e.g. unstable, runaway dynamics, with much faster decision times than the original ones.

For the LCA model, this approach seemingly proved most effective; SINDy was readily able to capture the decision variable dynamics (Figure 4C) in addition to choice behaviour for the model (Figure 3C). For the case of the NLB model, SINDy's predictive performance has also improved, especially with higher SNR (Figure 4D).

For the multi-trial approach, we followed the same data analysis as for the trial-averaged approach to allow for fair comparison. For DDM, the decision dynamics were well replicated, especially for higher SNRs, following a similar trend to the trial-averaged approach but with much improvement in terms of time course replication across SNRs (Figure 5A). The choice behaviour was drastically improved using multi-trial approach (Figure 3A), demonstrating its utility clearly for such a model. However, for LCA-DDM, it struggled to replicate the decision dynamics due to similar reasons mentioned in the trial-averaging approach above (Figure 5B). The choice behaviour using the best performing polynomial (order 0) still leaves something to be desired but it definitely offers an improvement on the trial-averaged approach's predictions. The LCA did perform well at replicating the decision dynamics (Figure 5C) and predicting choice behaviour but it is the only model that the multi-trial approach did not offer any improvement compared to the trial-averaged approach. For NLB the multi-trial approach improves on upon the trial-averaged method in terms of estimated decision dynamics (Figure 5D). The choice behaviour also improves overall but it does struggle to capture higher SNR choice accuracies, which could be due to the acceleration towards the steady state and leaving SINDy with less data to work with and an inaccurate model fit (Supplementary Figure S3).

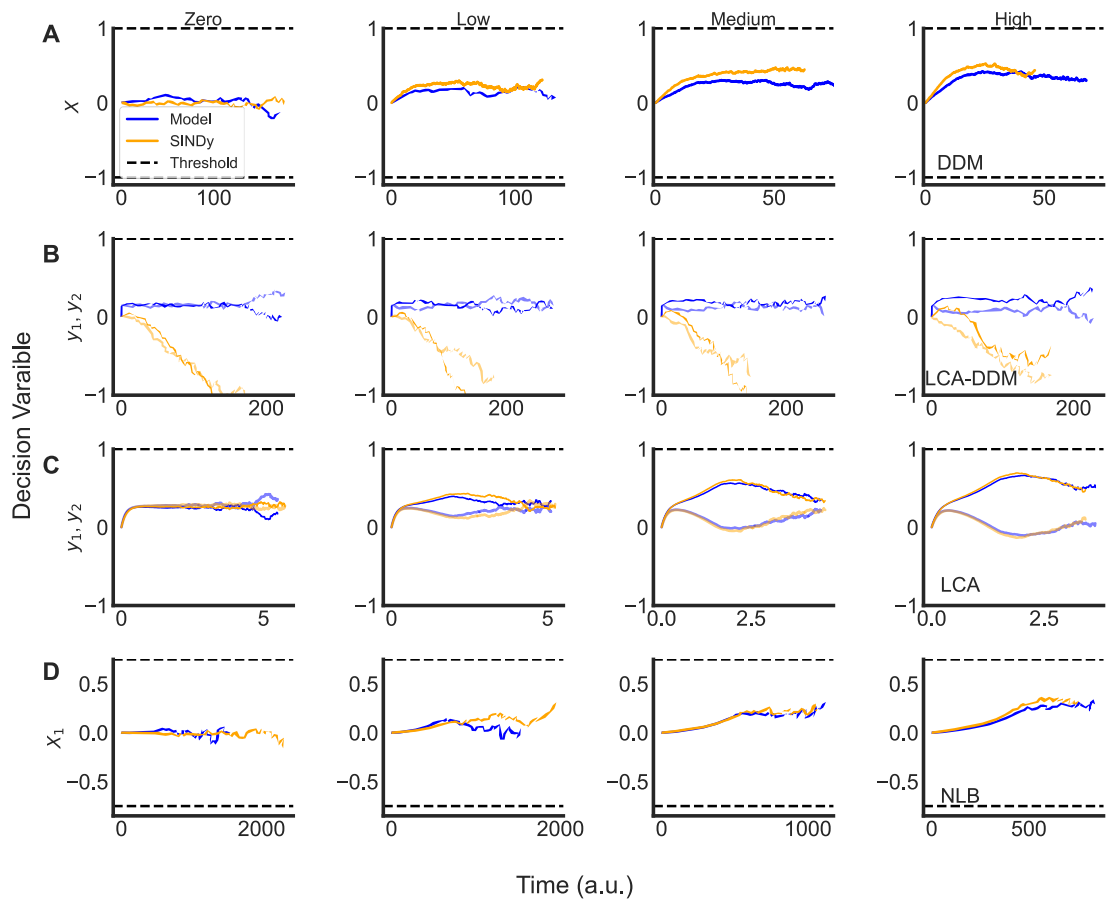

**Figure 4:** Trial-averaged of parameters approach's decision variable activities for different SNRs explain choice behavioural fits. Zero, low, medium, high SNRs; high SNR based on choice accuracy of both original and SINDy-derived models ≥ 95% and their differences being ≤ 1%. Low and medium SNRs determined from equally divided intervals between these two SNR values. Higher SNRs improved SINDY's model prediction. Labels as in Figure 2.

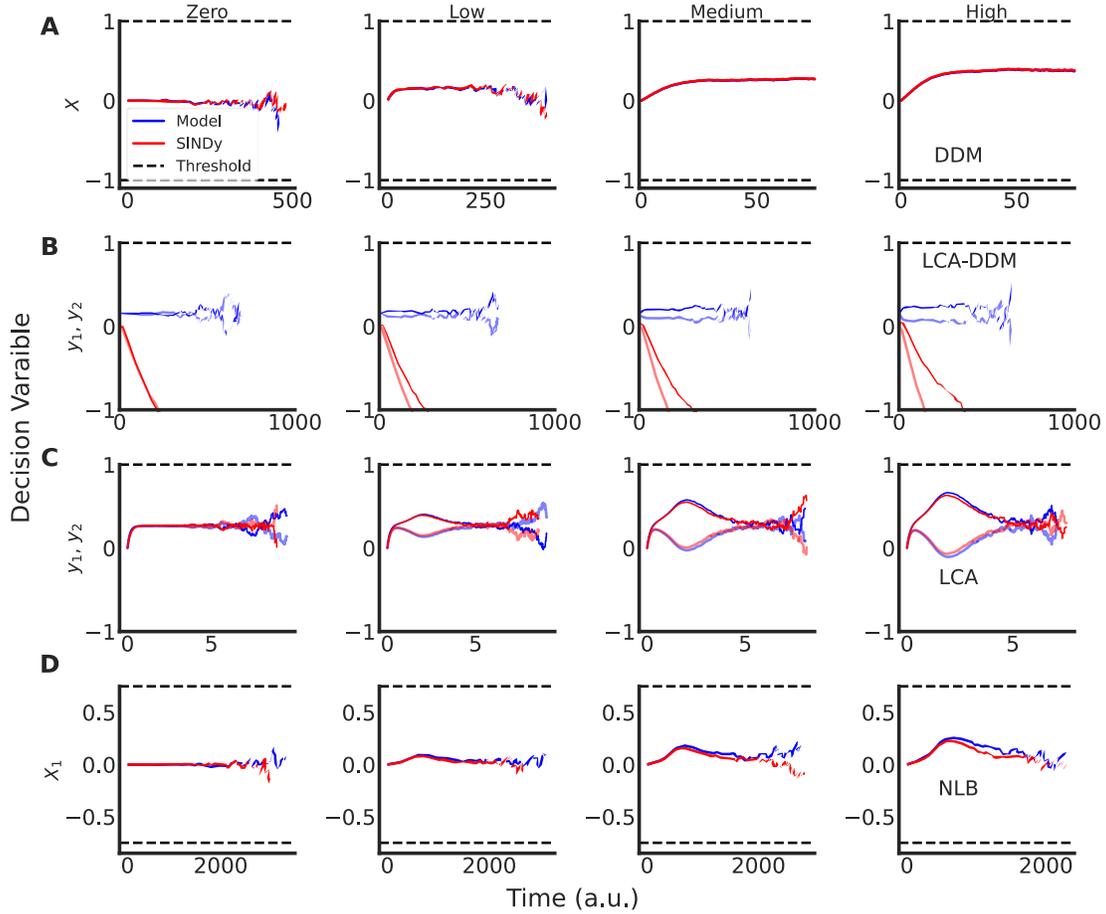

**Figure 5:** Multi-trial approach decision variable activities for different SNRs explain choice behavioural fits. Decision variable dynamics averaged over the same SNRs for comparison with those for the trial-averaged approach (compared with Figure 4). Blue (red): original (estimated) averaged decision variable activities.

Thus, with the trial-averaging approach, SINDy can generally handle the noisy dynamics better than single-trial SINDy approach and therefore predict choice behaviour with greater confidence compared to that of the single-trial approach.

The multi-trial SINDy further enhances SINDy's performance, with predicted choice behaviours and decision variable dynamics even closer to those of the original models. These results were also validated based on the differences between the model parameters (see Supplementary Notes 1, 2 and 3 for the highest SNR) and quantified by their RMSE values (Supplementary Figure S2).

## Discussion

Identifying the right decision-making model is important towards understanding underlying decision mechanisms. We have shown that the SINDy algorithm, especially utilising multiple trials, performs reasonably at recovering the underlying governing equations for the various often-used sequential sampling based decision-making models. Specifically, this was achieved by uncovering the dynamical equations from the DDM, linear and metastable LCA, its approximation to DDM (LCA -DDM), and the NLB model (Ratcliff, 1978; Usher & McClelland, 2001; Usher & McClelland, 2001; Bogacz et al., 2006; Zhou et al., 2009). We showed the potential of SINDy's utility replicating the deterministic portion of the decision variable profiles, which in turn aid in predicting choice behaviours. This not only advances our computational toolkit but also enriches our understanding of underlying decision-making processes, extending the foundational work by Brunton et al. (2016). Further, as the reaction-time task in the decision-making models is akin to first-passage time problems, wherein the systems' dynamics may only be partially observable, our results show that the application of SINDy may be extended to similar problems in other fields such as in engineering, physics, mathematics and biology (Gillespie, 1992; Kampen, 2011).

The adaptability of SINDy to predict choice accuracy and decision times across different SNR underscores its versatility. Moreover, SINDy can readily handle complex, nonlinear and multi-stable model (NLB) and finely tuned LCA-DDM. Additional spurious terms of higher orders for SINDy-derived NLB model may be minimised by considering smaller decision variable values, for instance closer to the beginning of trials.

Three different SINDy approaches were investigated in this study. The multi-trial SINDy approach demonstrated the best performance in elucidating choice behaviours and decision variable dynamics of all the original models. It also provided the fastest performance but was more computationally memory intensive (Supplementary Note 4). More modest performance was obtained using trial-averaging of model parameters across trials and doing especially well for the LCA model. This approach took much longer than the multi-trial approach but was less memory intensive. For the single-trial SINDy approach, although its performance was not as high as the above two methods, its utility might lie in modelling neural data in real-time (Raza et al., 2020). When compared with the other two approaches to fit both models using all trials at once, its memory requirements are less but its runtime is longer than the multi-trial approach (Supplementary Note 4). Hence, our work offers alternative SINDy approaches depending on the needs

Although the methods that used multiple trials were able to handle noise, the assumption of known noise characteristics may present some limitation, particularly when translating these models to analyse real neural data where noise information may not be readily available. This challenge is accentuated in LCA-DDM, where SINDy's performance varied, highlighting the algorithm's sensitivity to finely tuned parameters.

With some assumptions on the characteristics of the noise (e.g. additive and white noise like), perhaps more exact determination can be obtained through optimal fitting of choice behaviour of the original respective model, after the deterministic portion of the model is identified using the methods shown in this study. Future work should also evaluate SINDy on other nonlinear decision-making models (e.g. Marshall et al., 2022; Pirrone et al., 2022; Asadpour et al., 2024). Other similar data-driven methods could be explored and compared with SINDy (e.g. Pandarinath et al., 2018) within decision-making or first-passage time processes.

While applying SINDy to empirical neural data is beyond the scope of our investigation, neural correlates of evidence accumulation for decision-making, in the form of time series data, are well known and have been identified in invasive and non-invasive recordings across species (Hanks & Summerfield, 2017; O'Connell et al., 2018). For large-scale or brain-wide neural data, dimensional reduction may be required before further analysis or modelling (Cunningham & Yu, 2014). This is especially the case for decision-making which often resides in lower dimensional neural space (Ganguli et al., 2008; Shenoy & Kao, 2021; Steinemann et al., 2023). In fact, one of the advantages of SINDy is its ability to handle high-dimensional data (Brunton et al., 2016). Thus, this would be an interesting future direction to apply our developed methods. In conclusion, while we have successfully demonstrated SINDy algorithm's performance on cognitive computational models, in the process, its areas for improvement have also been revealed. For instance, enhancing the algorithm's handling of noise and its applicability to more complex models will be crucial for its successful integration into decision neuroscience research. Such advancements will not only provide deeper insights into the computational basis of decision-making but also improve our ability to interpret and predict neural activity underlying cognitive processes.

## Author Contributions

B.L. and K.W.-L. designed and conceptualised the analyses. B.L. performed the simulations and analyses. S.B. and K.W.-L. validated the codes, data and analyses. B.L., S.B. and K.W.-L. wrote the paper. K.W.-L. supervised the research.

## Acknowledgments

We thank Abdoreza Asadpour and Cian O'Donnell for useful discussions. B.L. was supported by Ulster University via Northern Ireland Department for the Economy (DfE). K.W.-L. was supported by HSC R&D (STL/5540/19) and MRC (MC_OC_20020). We are grateful for access to the Tier 2 High Performance Computing resources provided by the Northern Ireland High Performance Computing (NI-HPC) facility funded by the UK Engineering and Physical Sciences Research Council (EPSRC), Grant No. EP/T022175/1.

# Uncovering dynamical equations of stochastic decision models using data-driven SINDy algorithm

## Supplementary Information

Brendan Lenfesty[1], Saugat Bhattacharyya[1], KongFatt Wong-Lin[1,*]

[1]Intelligent Systems Research Centre, School of Computing, Engineering and Intelligent Systems, Ulster University, Magee campus, Derry~Londonderry, Northern Ireland, UK

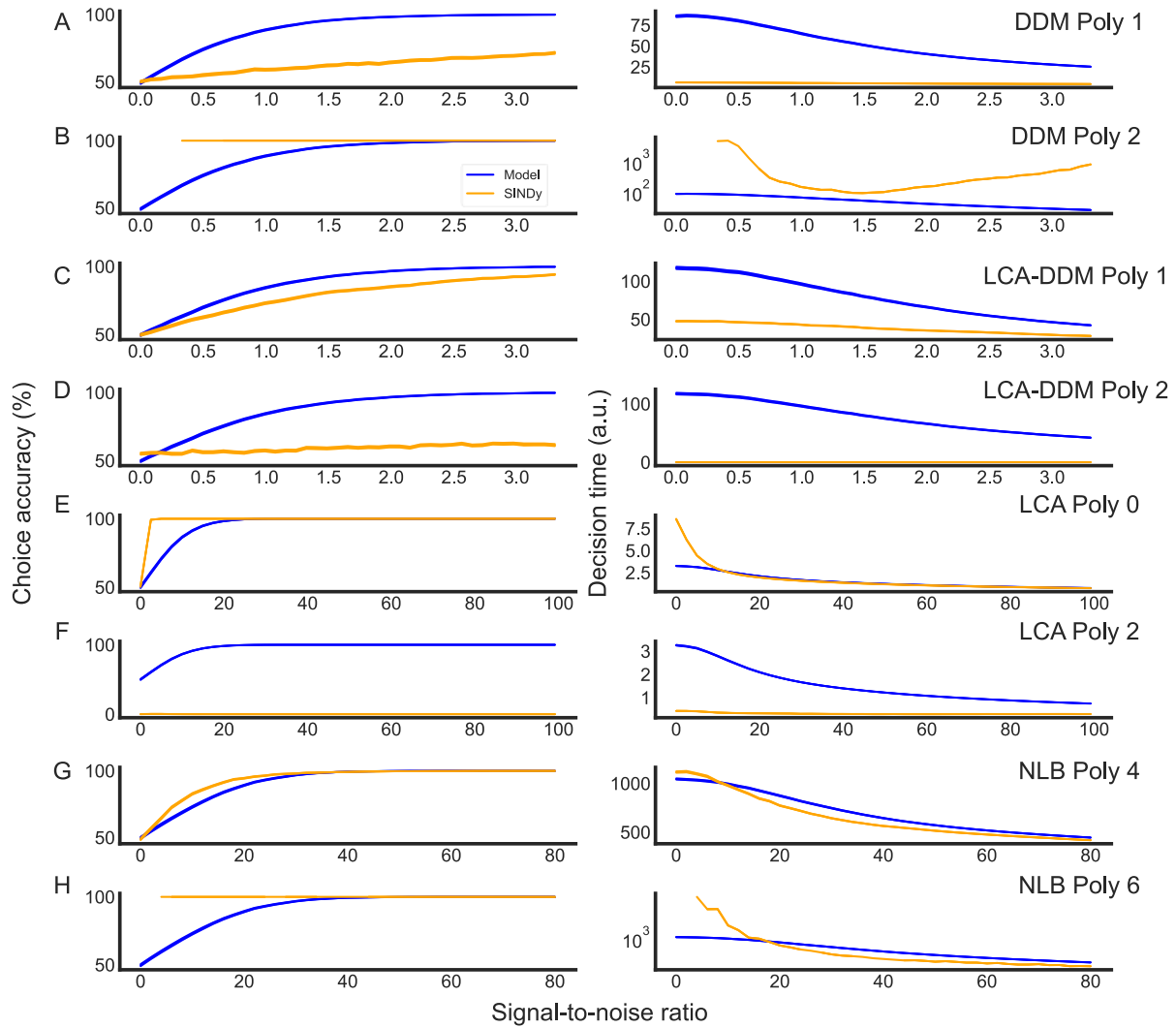

**Fig. S1**: Alternative polynomial orders investigated with trial-averaging of parameters approach. Refer to main manuscript (Figure 3, orange) for trial-averaged choice behaviour of the model parameters of the best performing polynomial order. Blue (orange): original (estimated) model's data. To determine the correct polynomial order for each decision model, we applied SINDy with a specific polynomial order, and averaged the parameter values obtained. We then compared the choice behaviours (accuracy and decision time) of each polynomial order with that of the original decision model to elucidate the best performing system of equations to predict the choice behavioural data. (B,H) For SINDy models of DDM polynomial order 2 and NLB polynomial order 6 at lower signal-to-noise ratios, the maximum simulation time was reached and no decision was made on these trials due to an issue with the SINDy models where the coefficient(s) uncovered for the model do not allow for the decision threshold to be reached, and thus resulting in a non-decision trial.

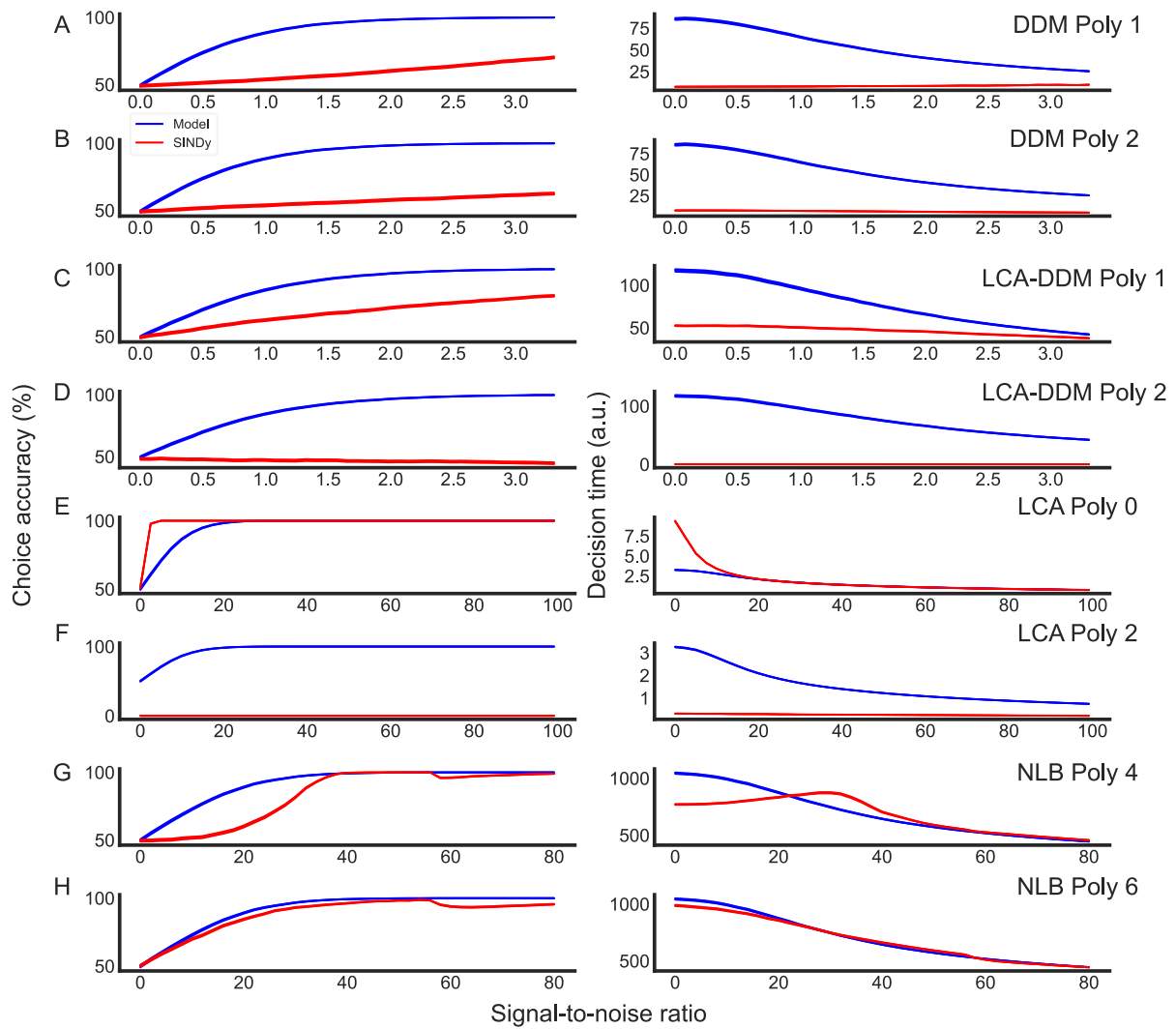

**Fig. S2**: Alternative polynomial orders investigated based on multi-trial approach. Refer to main manuscript (Figure 3, red) for multi-trial choice behaviour for the best performing polynomial order. Blue (red): original (estimated) model's data. To determine the correct polynomial order for each decision model, we applied SINDy with a specific polynomial order to fit to all trials for a given signal-to-noise ratio. We then compared the choice behaviours (accuracy and decision time) of each polynomial order with that of the original decision model to elucidate the best performing system of equations to predict choice behaviour.

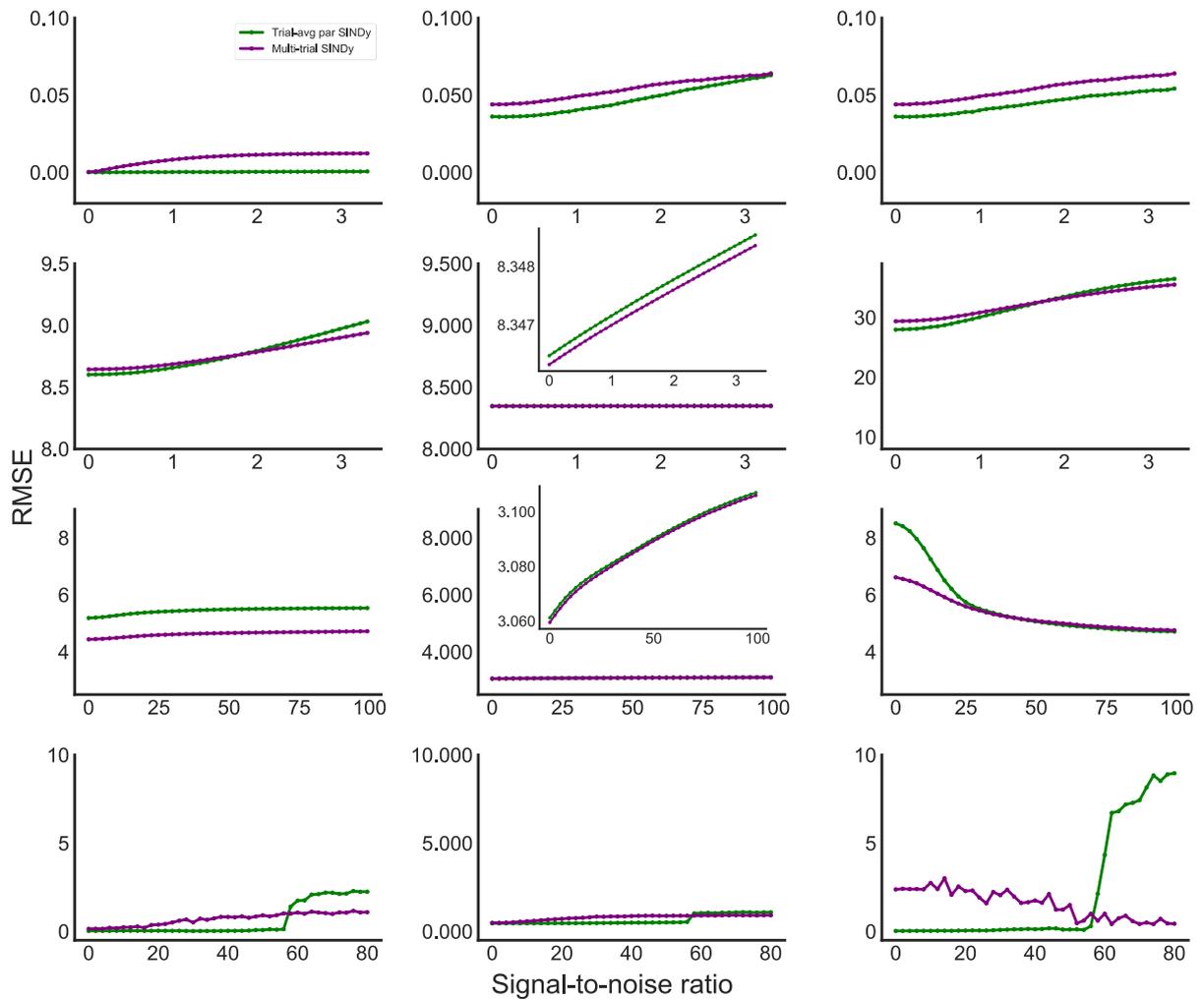

**Fig. S3:** Root mean squared error (RMSE) comparison between parameter trial-averaging and multi-trial approaches for simulations of each model for all polynomial orders. Columns from top to bottom: DDM, LCA-DDM, LCA and NLB. The polynomial orders for each row are as follows, from left to right: DDM estimated model polynomial 0, 1 and 2; LCA-DDM estimated model polynomial 1, 0 and 2; LCA estimated model polynomial 1, 0 and 2; and NLB estimated model polynomial 5, 4 and 6. Purple (green): trial-averaged (multi-trial) approach's RMSE. Inset: Zoomed in.

**Supplementary Note 1**: Sample single-trial estimations by SINDy based on correct polynomial order of model fitting.

DDM: SINDy estimated the drift-rate parameter to be $0.043$ while the original drift rate was $0.04$.

LCA-DDM: SINDy estimated the decay rates of the two mutually inhibiting decision units to be $k_1 = -1.51$ and $k_2 = -0.95$, their mutual inhibitory coupling constants to be $b_1 = -1.46$ and $b_2 = -1.0$ and their stimulus inputs to be $S_1 = 0.51$ and $S_2 = 0.3$. The corresponding parameter values of $k_1, k_2 = -10$ and $b_1, b_2 = -10$ and $S_1 = 3.04$ and $S_2 = 3$, and are significantly different from that of the SINDy-derived model, especially for the $k's$ and $b's$.

LCA: SINDy's derived model of the linear LCA decay rates of the two mutually inhibiting decision units were $k_1 = -1.99$ and $k_2 = -1.19$, their mutual inhibitory coupling constants were $b_1 = -2.64$ and $b_2 = -3.15$, and the stimulus inputs to be $S_1 = 2.6$ and $S_2 = 1.3$. Instead, the original values were $k_1, k_2 = -3$ and $b_1, b_2 = -4$ and $S_1 = 3.05$ and $S_2 = 1.85$ for the linear LCA.

NLB: NLB original model parameters were $b = 0.008$ and $\varepsilon = 0.05$ with coefficients of $-1$ in the cubic $X^3$ and quintic $X^5$ polynomials, while the SINDy-derived model parameter values were $b = 0.02$ and $\varepsilon = -0.17$, with coefficients of $-4.03$ ($X^3$) and $-5.4$ ($X^5$), and also came along with additional terms of $1.51 X^2$ and $7.75 X^4$.

**Supplementary Note 2**: Sample trial-averaged of parameters estimations by SINDy based on correct polynomial order of model fitting.

DDM: SINDy estimated the drift rate was 0.052 while the original value was 0.04.

LCA-DDM: The derived model had parameters $k_1 = -1.12$, $k_2 = -1.11$, $b_1 = -1.14$, $b_2 = -1.18$, $S_1 = 0.38$ and $S_2 = 0.34$, while the original values were $k_1, k_2 = -10$, $b_1, b_2 = -10$, $S_1 = 3.04$ and $S_2 = 3$.

LCA: The SINDy-derived LCA model had parameter values of $k_1 = -3.01$, $k_2 = -2.92$, $b_1 = -4.10$, $b_2 = -4.07$, $S_1 = 3.09$ and $S_2 = 1.9$, while the original values were $k_1, k_2 = -3$, $b_1, b_2 = -4$, $S_1 = 3.05$ and $S_2 = 1.85$.

NLB: For the NLB model, the original parameter values were $b = 0.008$ and $\varepsilon = 0.05$, with coefficients of $-1$ for the $X^3$ and $X^5$ terms, while the SINDy-derived NLB model parameters were $b = 0.009$ and $\varepsilon = 0.072$, with coefficients of $2.180\,(X^3)$ and $0.231\,(X^5)$, and additional terms $-0.279\,X^2$ and $-2.037X^4$.

**Supplementary Note 3**: Sample multi-trial estimations by SINDy based on correct polynomial order of model fitting.

DDM: SINDy estimated the drift rate was 0.041 while the original value was 0.04.

LCA-DDM: The derived model had parameters $k_1 = -0.87$, $k_2 = -0.87$, $b_1 = -0.9$, $b_2 = -0.92$, $S_1 = 0.28$ and $S_2 = 0.27$, while the original values were $k_1, k_2 = -10$, $b_1, b_2 = -10$, $S_1 = 3.04$ and $S_2 = 3$.

LCA: The SINDy-derived LCA model had parameter values of $k_1 = -2.89$, $k_2 = -2.3$, $b_1 = -3.79$, $b_2 = -3.78$, $S_1 = 3.0$ and $S_2 = 1.75$, while the original values were $k_1, k_2 = -3$, $b_1, b_2 = -4$, $S_1 = 3.05$ and $S_2 = 1.85$.

NLB: For the NLB model, the original parameter values were $b = 0.008$ and $\varepsilon = 0.05$, with coefficients of $-1$ for the $X^3$ and $X^5$ terms, while the SINDy-derived NLB model parameters were $b = 0.006$ and $\varepsilon = 0.13$, with coefficients of 3.68 ($X^3$) and 1.3 ($X^5$), and additional terms $-0.744\,X^2$ and $-4.16\,X^4$.

**Supplementary Note 4**: Computational time and (peak) memory costs for each of the three SINDy approaches' parameter fitting over all trials for all SNRs.

- Single-trial approach – computational time: 8866.03 seconds; peak memory: 110059.80MB
- Trial-averaged of parameters approach – computational time: 8798.85 seconds; peak memory: 110056.72MB
- Multi-trial approach – computational time: 3536.91 seconds; peak memory: 111688.21MB